\documentclass[letter]{aa} 

\usepackage{graphicx}
\usepackage{txfonts}
\usepackage{epstopdf}
\usepackage{subfig}
\usepackage{longtable}
\usepackage{appendix}
\usepackage[utf8]{inputenc}

\begin{document} 

\title{Direct evidence of a full dipole flip during the magnetic cycle of a sun-like star}
   \author{S. Boro Saikia
          \inst{1}, T. Lueftinger \inst{1}, S. V. Jeffers \inst{2}, C. P. Folsom \inst{3,}\inst{4},V. See \inst{5}, P. Petit \inst{3,}\inst{4}, 
          S. C. Marsden \inst{6}, A. A. Vidotto \inst{7}, J. Morin \inst{8}, A. Reiners \inst{2}, M. Guedel \inst{1}, and the BCool collaboration}

   \institute{University of Vienna, Department of Astrophysics,
              T\"urkenschanzstrasse 17, 1180 Vienna, Austria \label{inst1} \and
              University of Goettingen, Institute for Astrophysics,
              Friedrich Hund Platz 1, 37077, Goettingen, Germany \label{inst2} \and 
              Universit\'e de Toulouse, UPS-OMP, IRAP, Toulouse, France \label{inst3} \and
              CNRS, Institut de Recherche en Astrophysique et Planetologie, 14, avenue Edouard Belin, F-31400 Toulouse, France \label{inst4} \and
            University of Exeter, Department of Physics $\&$ Astronomy, Stocker Road, Devon, Exeter, EX4 4QL, UK \label{inst5} \and
            University of Southern Queensland,Centre for Astrophysics, Toowoomba, QLD 4350, Australia \label{6} \and
            School of Physics, Trinity College Dublin, University of Dublin, Dublin-2, Ireland \and
            LUPM-UMR 5299, CNRS $\&$ Universit\'e Montpellier, place Eug\`ene Bataillon, 34095 Montpellier Cedex 05, France}
\titlerunning{Dipole}
\authorrunning{S.Boro Saikia et al. }
   \date{}

 
  \abstract
   {The behaviour of the large-scale dipolar field, during a star's magnetic cycle, 
can provide valuable insight into the stellar dynamo and associated magnetic field manifestations such as stellar winds.} 
   {We  investigate the temporal evolution of the dipolar field of the K dwarf 61 Cyg A using spectropolarimetric observations covering  nearly one 
magnetic cycle equivalent to two chromospheric activity cycles.}
   {The large-scale 
magnetic field geometry is reconstructed using Zeeman Doppler imaging, a tomographic inversion technique. Additionally, the chromospheric activity
is also monitored.}
{The observations provide an unprecedented sampling of the large-scale field over a single magnetic cycle of a star other than the Sun. Our results 
show  that 61 Cyg A has a dominant dipolar geometry except at chromospheric activity maximum.
The dipole axis migrates from the southern to the northern hemisphere during the magnetic cycle. It is located at higher latitudes at chromospheric activity 
cycle minimum and at middle latitudes during cycle maximum. The dipole is strongest at activity cycle minimum and much weaker at activity cycle maximum.}
{The behaviour of the large-scale dipolar field during the magnetic cycle resembles the solar magnetic cycle. Our results are further confirmation 
that 61 Cyg A indeed has a 
large-scale magnetic geometry that is comparable to the Sun's, despite being a slightly older and cooler K dwarf.}
\maketitle
\section{Introduction}
In recent years, with the advances made in stellar magnetic field observations, the technique of 
Zeeman Doppler imaging (ZDI) \citep{semel89,brown91,donatibrown,piskunov02,kochukhov02} 
has been able to reveal the diversity in stellar large-scale magnetic field geometry. 
Both single-epoch studies (Petit et al. in prep.) and long-term monitoring \citep{fares09,morgenthaler12,
jeffers14,borosaikia15,jeffers17,fares17} have revealed that stellar 
magnetic geometries do not necessarily have the same large-scale structure as the Sun. While older slowly rotating stars exhibit a more sun-like, 
almost purely poloidal large-scale magnetic field \citep{petit08}, younger rapidly 
rotating stars can have magnetic geometries that do not bear resemblance to the solar large-scale field 
\citep{morgenthaler12,borosaikia15, rosen16,hackman16}. Surprisingly, until now, 
sun-like magnetic geometry evolution has been an anomaly, not the expected norm\footnote{A solar-type star has a sun-like field evolution 
when the large-scale magnetic field flips polarity in sync with its activity cycle. Additionally, the large-scale field  changes field 
complexity during the activity cycle in  the same way as the Sun does.}. 
To date, approximately 20  stars have been observed over multi-epochs, the majority of which 
were observed by the BCool collaboration \footnote{https://bcool.irap.omp.eu/} \citep{marsden14}. Of these stars only a few 
exhibit polarity flips \citep{petit09,morgenthaler11,fares09,mengel16,borosaikia16,jeffers18}, and only 61 Cyg A \citep{borosaikia16} and $\tau$ Boo 
\citep{jeffers18} are known to exhibit polarity flips of the large-scale field in phase with the star's chromospheric activity cycle.
However, no star has been observed over a full magnetic cycle (or two consecutive activity cycles); 
one solar magnetic cycle is the time taken by the polarity of the large-scale field to switch to an opposite polarity
and flip back to the original sign. 
This work, for the first time,  investigates the evolution of the large-scale field of a sun-like star over its magnetic cycle.  
\paragraph{}
The K dwarf 61 Cyg A was observed for a full activity cycle or half a magnetic cycle in paper I \citep{borosaikia16}. ZDI reconstructions 
of this star have revealed a magnetic field geometry that flips its polarity in sync with its chromospheric activity cycle. 
The field geometry is dominantly poloidal. During activity minimum the poloidal field is strongly dipolar, and at activity 
maximum quadrupolar and octupolar fields dominate. 
In the Sun, the dipolar field is known to undergo spatial and temporal evolution during the solar magnetic cycle \citep{sanderson03,derosa12,vidotto18}. 
The evolution of the 61 Cyg A large-scale field is similar 
to the evolution of the solar large-scale field over its activity cycle. 
However, the star was observed
for only one activity cycle in paper I and it could not be established whether such evolution is repeated periodically. 
If 61 Cyg A has a sun-like dynamo process operating like that of  the present-day Sun, 
then the strength and location of the dipolar field should have a sun-like
evolution during the full magnetic cycle. 
In this work, the large-scale field of 61 Cyg A is monitored, with strong emphasis on the dipole field, 
for almost a full magnetic cycle by combining new multi-epoch observations (activity cycle II) with epochs from paper I (activity cycle I). 
\section{Observations}
Simultaneous Stokes $V$ (circularly polarised) and Stokes $I$ (unpolarised) spectra were taken  
using the high-resolution spectropolarimeter NARVAL at Telescope Bernard Lyot (TBL), Pic du Midi 
\citep{auriere03}. 
Six epochs of observations were obtained spanning three years (2015.77, 2015.91, 
2016.50, 2017.50, 2017.89, 2018.52). Epochs 2017.50, 2017.89, and 2018.52 
were observed as part of the BCool collaboration. We also use data published in paper I, 
where six epochs were observed over nine years. Furthermore, epoch 2005.47 
was observed as part of POLARBASE \citep{petit14}. However, care should be taken when 
interpreting the 2005.47 results as the rotational phase coverage is sparse. 
\paragraph{}
To detect Zeeman signatures in 
Stokes $V$, the multi-line technique least-squares deconvolution (LSD) \citep{donati97,kochukhov10} is applied, which boosts the signal-to-noise ratio. 
The LSD Stokes $V$ and Stokes $I$ profiles are obtained following the same procedure as applied in paper I. 

\section{Methods}
Zeeman Doppler imaging is the only
technique that can reconstruct the large-scale surface magnetic geometry in sun-like stars.  It is a 
tomographic technique that inverts LSD Stokes $V$ profiles to reconstruct the magnetic maps. 
The code iteratively fits synthetic Stokes profiles obtained from a 
stellar model to observed 
Stokes spectra, 
until the desired reduced $\chi^2$ is reached \citep[see][ and the references therein for more details on the ZDI technique]{donati06,folsom16}. 
Since it is an inverse problem, the maximum entropy technique is utilised as a regularisation tool \citep{skilling84}. 
\paragraph{}
The stellar parameters (inclination = 70 $^\circ$, $v \sin i$ = 0.92 kms$^{-1}$) 
used to reconstruct the magnetic maps are taken from paper I. A rotation period of 34.2 days and HJD 2454308.49809 as zero phase are used 
to calculate the phases (taken from paper I). The maximum spherical harmonic degree $l_\mathrm{max}$ of 11 is 
applied to all of our maps.\footnote{A $l_\mathrm{max}$ value of 11 was also applied in paper I. As the majority of the magnetic energy is in the lower order 
components, $l_\mathrm{max}\geq$11 does not improve the quality of the maps.} 
The differential rotation parameter determined in paper I is 
applied here. 
To determine differential rotation ZDI requires a dense phase coverage \citep{morgenthaler12} and a complex field 
geometry so that the code can track different magnetic features as the star rotates and determine the surface shear. 
The field geometry of 61 Cyg A is predominantly dipolar with single
polarity in the observed hemisphere. Since there is no change in surface geometry in the visible hemisphere it is extremely hard to converge to the correct 
differential rotation. We have thus used the differential
rotation from activity cycle I (paper I). Using the differential rotation 
parameters from activity cycle I, a 
reduced $\chi^2$ of 1.0 is achieved for all epochs except 2016.50, where the reduced $\chi^2$ is 1.2. 
\begin{figure}
\centering
\includegraphics[width=1.1\columnwidth]{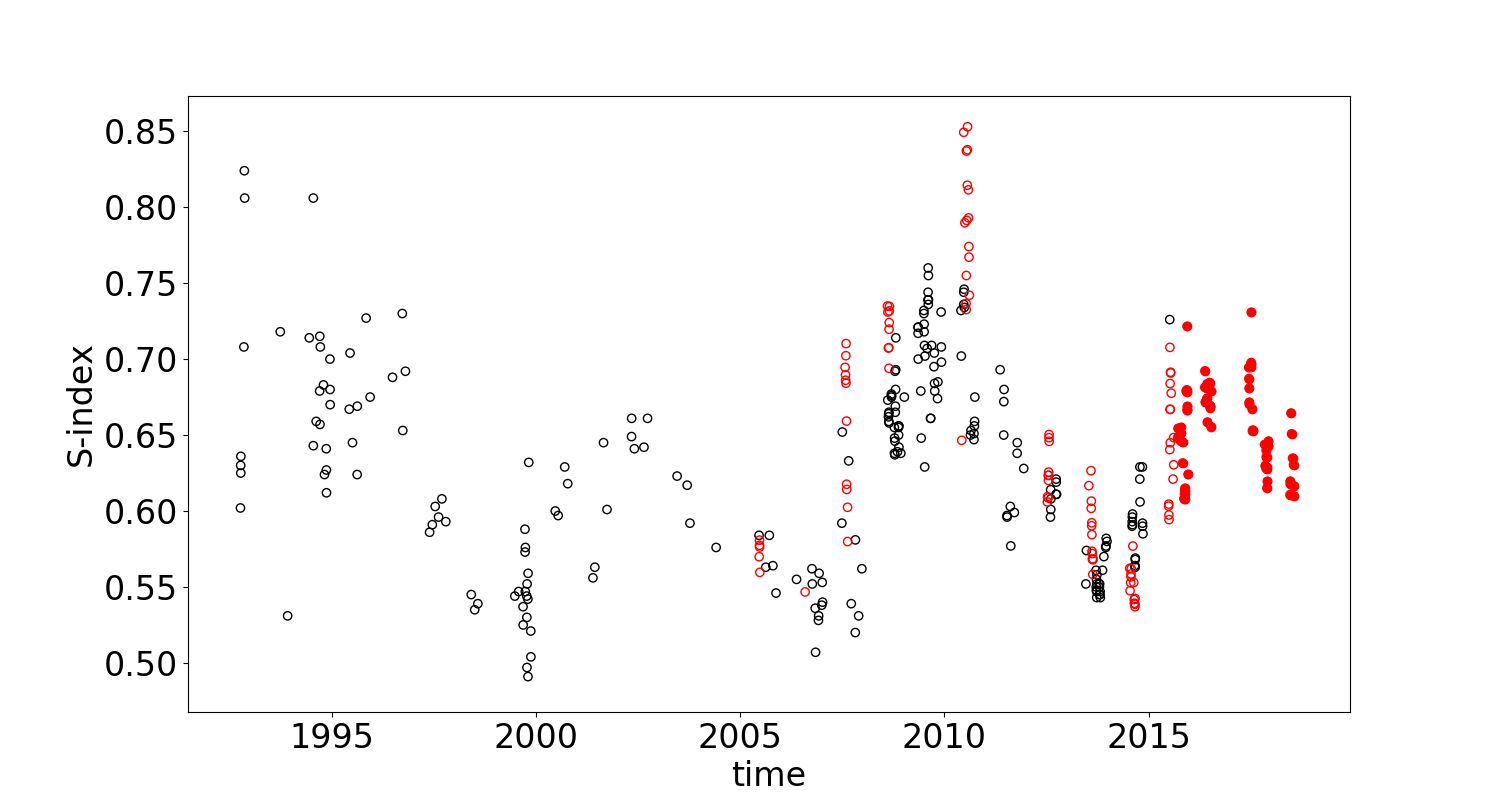}
\caption{S-index of 61 Cyg A measured using data from NARVAL/TBL and Lowell observatory (taken from paper I) together with new measurements from this 
work. The Lowell data is shown as black open circles and the old NARVAL/TBL data is shown as red open circles. The red filled circles are data from this
work.}
\label{appendixsindex}
\end{figure}

\section{Results}
\subsection{Chromospheric activity cycle II}
The chromospheric activity (S-index) is determined using the same method as in paper I. Figure \ref{appendixsindex} shows that 
61 Cyg A has a periodic activity cycle of varying amplitude. The current cycle has a lower amplitude compared to cycle I. 
The length of cycle II also appears to be shorter than activity cycle I. However, cycle II is not over and new 
observations will provide a better estimate of its length. 
\subsection{Magnetic geometry}
\begin{figure}
\centering
\includegraphics[width=1.\columnwidth]{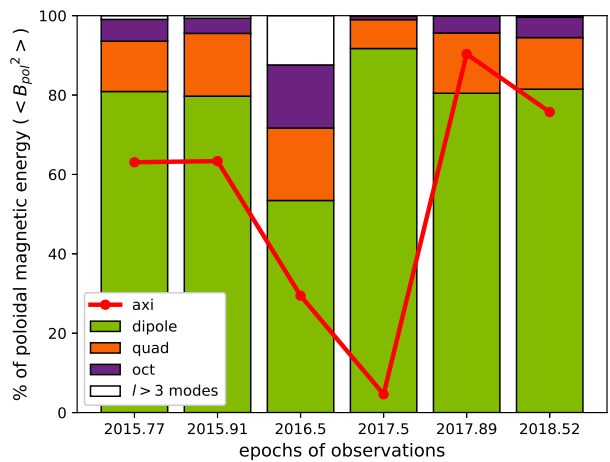}
\caption{Magnetic energy distribution in different components of the 61 Cyg A poloidal field during the current activity cycle (cycle II).
Magnetic energy in the dipolar component is shown in green, the quadrupolar 
component is shown in orange, the octupolar component is shown in purple, and  higher order components ($l>$3) are shown in white. The red line shows the 
axisymmetry of the total field.}
\label{energy}
\end{figure}

ZDI maps of 61 Cyg A in cycle II (this work) are dominantly poloidal. The poloidal field is concentrated at 
lower order spherical harmonics: dipolar, quadrupolar, and octupolar, as shown in Fig. \ref{energy} 
(Fig. \ref{polappendix} shows the poloidal magnetic energy distribution over the magnetic cycle of 61 Cyg A, epoch 2005.47 to  2018.52). 
More than 80$\%$ of the poloidal energy is dipolar except in 2016.50, close to activity maximum in cycle II (Table \ref{comps}).  
\paragraph{\textbf{From activity minimum to maximum:}}
During the first two epochs (2015.77, 2015.91) the activity increases and moves towards activity maximum.  
During this period the radial field exhibits a strong positive polarity dipole in the observed hemisphere, as shown in Fig. \ref{A1}. 
The azimuthal field shows a strong band of negative polarity at equatorial latitudes in 2015.77, which gets weaker in 2015.91. 
The meridional field shows a positive
polarity band also at equatorial latitudes that remains consistent.
\paragraph{\textbf{Near activity maximum}}
During epoch 2016.50 the large-scale field appears to convert 
from a simple dipolar geometry (Fig. \ref{A1}). The azimuthal field gets stronger and the 
dipolar energy gets weaker at this epoch. A small fraction of the magnetic energy is detected in $l_\mathrm{max}>$3. The higher order modes were also
detected at activity maximum during activity cycle I.

\paragraph{}  
At epoch 2017.50, the radial field is reconstructed with both strong negative and positive regions (Fig. \ref{A2}),
indicating a dipole whose axis is aligned to the equator. The meridional and azimuthal fields are considerably weaker. The field is almost purely poloidal 
and the dipole energy is at its strongest. At 2017.50 the negative polarity of the dipole gets stronger, suggesting a change in the dominating polarity in
the coming epochs.
\paragraph{\textbf{From activity maximum to minimum}}
During epochs 2017.89 and 2018.52 the radial field switches polarity; with the appearance of a strong negative dipolar field at the visible hemisphere 
(Fig. \ref{A2}). The meridional field also 
exhibits a negative polarity field, but at equatorial latitudes. 
The azimuthal field appears as a more complex field in 2018.52 
(see Appendix A for a more detailed description on the field geometry).

\begin{figure}
\centering
\includegraphics[width=1\columnwidth]{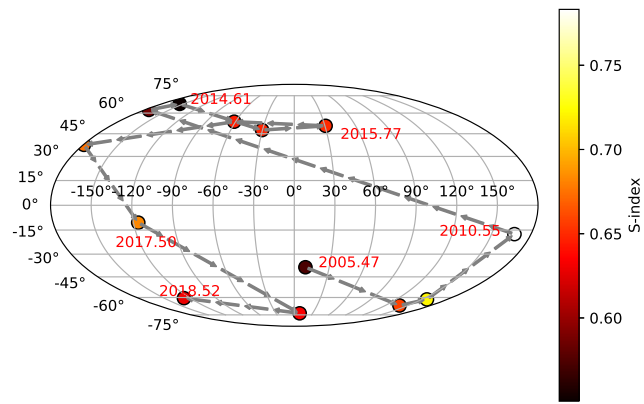}
\caption{Location of the dipole axis during activity cycle I and II ($\sim$1 magnetic cycle), starting at the lower right and moving  anti-clockwise. 
All values 
except 2005 are taken from this work or paper I. The colour bar 
represents the chromospheric activity index S-index. The arrow shows the direction of the dipole axis movement from 2005.47 to 2018.52. Some 
representative dates are  shown in red.} 
\label{dipolelocation}
\end{figure}
\begin{figure*}
\centering
\includegraphics[width=1\linewidth]{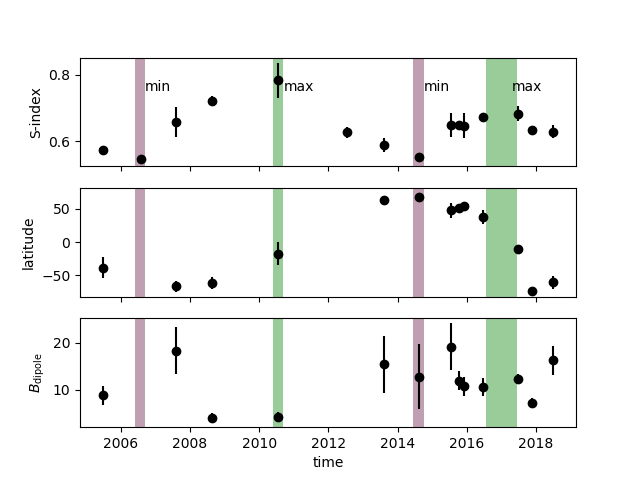}
\caption{Chromospheric activity (S-index, top) cycle, dipole location (middle), and strength (bottom) over time. \textit{Top}: S-index cycle, \textit{middle}: Dipole location
 and \textit{bottom}: Dipole strength at the axis as a function of time. The standard deviation is shown as an error bar.  Activity minimum is shown in purple and  activity maximum is shown in green.}
\label{main}
\end{figure*}
\subsection{Dipole migration}
One key observation from  Figs. \ref{A1} and \ref{A2} is the changing inclination of the radial dipolar field. 
In order to quantify the inclination of 
the dipole we calculated the location of the dipole axis for all available epochs,  starting with 2005.47 up to 2018.52. 
The axis of the dipole is obtained by finding the maximum of the 
dipole field (positive pole). Figure \ref{dipolelocation} 
shows the location 
of the dipolar axis, where it moves from one hemisphere to the other and comes back to the original hemisphere during a full magnetic cycle.
This evolving nature of the dipolar axis location is a common feature of the solar large-scale field 
\citep{derosa12}. 
\subsection{Dipolar field and stellar activity}
\begin{figure}
\centering
\includegraphics[width=1\columnwidth]{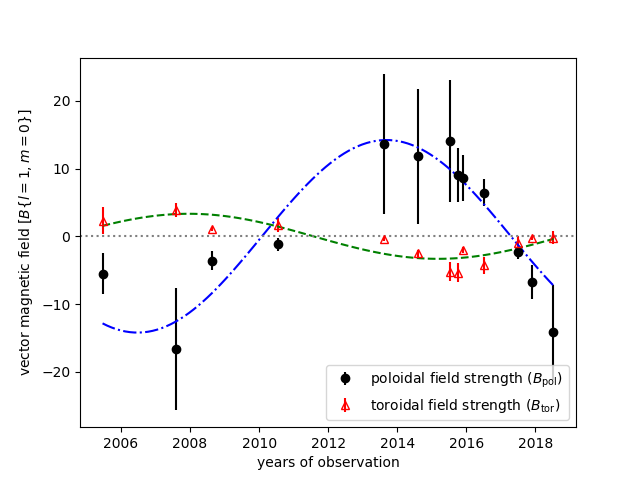}
\caption{Axisymmetric dipole field strength, $B (l=1, m=0)$, both poloidal (black filled  circles) and toroidal (red open triangles) 
components of the magnetic field. The dotted horizontal line in grey represents $B$=0. The sinusoidal 
fits are obtained with a magnetic cycle period of 14.4 years.
The error bars are dispersion obtained in a  way similar to that in paper I. This is a continuation of Fig. 11 in paper I.}
\label{fig11}
\end{figure}

Figure \ref{main} shows the evolution of the dipole latitude and strength in conjunction with the S-index cycle. The data for cycle I is taken from paper I.
Mean S-indices are shown for each epoch. 
The standard deviations are shown as error bars. Cycle minima are shown as pink vertical bars and cycle maxima as green bars. 
As the exact occurrence of  activity maximum is not known, it could be anytime in the one-year period around
the second green vertical bar in Fig. \ref{main}. 
\paragraph{}
In Fig. \ref{main} we are missing two data points in the bottom two subplots, 2006 and 2012. 
Only one spectropolarimetric observation was taken in 2006, first minimum, 
which is not enough to reconstruct 
a magnetic map. The spectropolarimetric observations in 2012 had polarisation anomalies (See paper I for more details); as a result,  
the magnetic map could not be reconstructed. 
As shown in Fig. \ref{main}, the dipolar axis is located near polar latitudes at activity minimum and at equatorial latitudes at activity
maximum. Although we do not have any data points at first minimum and second maximum (Fig. \ref{main} middle), the overall trend in dipole location 
shows that there is a strong correlation with activity cycle phase, and both exhibit periodic behaviour. The dipole axis evolution agrees with a magnetic
cycle period of $\sim$ 14 years.
\paragraph{}
The strength of the dipole, $B_\mathrm{dipole}$, is also  plotted against time, as shown in the bottom subplot in Fig.\ref{main}. 
The maximum of the dipole field, which occurs at the dipole axis, is taken to be the dipole strength in
Fig. \ref{main}. We do not have a magnetic map for 
the first minimum in cycle I; 2007 and 2005 are the closest to the first minimum in activity cycle I. 
The dipolar field is at its weakest at the 
first maximum. During the second minimum the dipole increases in strength. 
Although the peak of the second maximum is not observed, close to the second 
activity  maximum the dipole field remains relatively strong. No correlation is detected between the dipole field strength and the magnetic cycle.
\paragraph{}
Figure \ref{fig11} shows the vector field strength of the axisymmetric dipole ($l$=1, $m$=0) in both poloidal and toroidal components 
(similar to Fig. 11 in paper I). The poloidal axisymmetric dipole is calculated by taking the maximum of the vector 
$B (l=1,m=0)$ at the visible rotational pole. The 
toroidal axisymmetric dipole is calculated by measuring the maximum signed $B (l=1,m=0)$ at the nearest toroidal band to the rotational pole. 
 The poloidal dipole exhibits cyclic variation with a period that correlates to the dipole axis location in 
Fig. \ref{main}. The sign
of the poloidal dipole field changes with the magnetic cycle phase. 
The toroidal dipole exhibits a weak anti-correlation with the poloidal dipole field. The errors in Figs. \ref{main} (middle and bottom panel) 
and  \ref{fig11} are determined via 
the same method used in paper I (see Section 5.1).
\subsection{Evidence of the dipole flip in LSD Stokes $V$ profiles}
The dipole flip of 61 Cyg A is also seen directly in the LSD Stokes $V$ profiles. 
Figure \ref{sv} shows the LSD Stokes $V$ profiles from epochs 2015.77 to 2018.52 in the same scale. The Stokes $V$ profiles
are shifted and colour-scaled to their rotational phase. Although the observed phases are not uniform in 
every single epoch, there are multiple observations with similar phases enabling direct comparison between the epochs. During epochs of single dominant 
polarity at the observed hemisphere (in the ZDI maps), 
the dominant Stokes $V$ profiles all share the same sign, with a positive lobe on 
one side of the line and a negative lobe on the other, although the amplitude may vary.
During epochs of mixed polarity on the visible hemisphere, the Stokes $V$ profiles exhibit both positive and negative signs. 
The epochs 2015.77 and 2015.91 show dominant positive Stokes $V$ profiles, whereas epochs 2016.50 and 2017.50 show both positive and negative profiles. 
Finally epoch 2017.89 and 2018.52 show dominant negative Stokes $V$ profiles. This change in sign 
provides  model independent evidence of the large-scale field polarity reversal (see Appendix B).    
\section{Discussion and conclusions}
The observations taken in this work in combination with archival
data (paper I + \citealt{petit14}) result in an observational time span of $\sim$14 years, providing a unique opportunity to monitor the 
large-scale surface magnetic geometry of 61 Cyg A over its magnetic cycle (two simultaneous activity cycles). 
The current activity cycle has a smaller amplitude and length than activity cycle I (paper I). 
This shows that although the cycle is quasi-periodic, 
it is not uniform in length. Previous observations of the star's cycle have also shown this non-uniformity in amplitude and 
period length \citep{baliunas95} (also see 
paper I), although the period length does not vary dramatically for a short amplitude cycle. Further observations of the current
cycle are required to determine the true cycle period of cycle II.  The solar cycle is also known to vary in length \citep{hathaway10}.
\paragraph{}
The large-scale field of the star is strongly poloidal throughout cycle II. 
The poloidal field is constituted of mostly lower order spherical harmonics,  of which the dipole is the
strongest. The dipole remains consistently dominant throughout the observations except close to activity maximum. 
We do not have observations at the peak of maximum during
cycle II, yet the dipole field gets considerably weaker during epoch 2016.50. At epoch 2016.50 the quadrupolar, octupolar, and 
higher order components have  higher energies compared to other epochs. Similar 
weakening of the dipole was also detected during activity maximum in cycle I.
\paragraph{}
Similar  to the solar case, the large-scale dipole 
flips its polarity in sync with the chromospheric 
activity cycle in a quasi-periodic manner. The dipole field reverts  to its original polarity during our combined $\sim$ 14 years of observation. 
The dipole axis location has a tight correlation
with the star's magnetic cycle. The axisymmetric vector dipole field also 
evolves together with magnetic cycle indicating a period of $\sim$ 14 years.
Overall our results show that the evolution of the 61 Cyg A large-scale field, specifically the dipole field, 
is very similar to the solar case indicating common dynamo processes. 
The solar magnetic cycle, in the large-scale, could be a template for sun-like stars like 61 Cyg A and vice versa. 
\begin{acknowledgements}
We thank Matthew Mengel for providing us with some of the data and the TBL team for their help during the observations. 
S.B.S and T.L. acknowledge funding via the Austrian Space Application Programme (ASAP) of the Austrian Research Promotion Agency (FFG) within ASAP11, the 
FWF NFN project S11601-N16, and the sub-project S11604-N16. S.V.J. acknowledges the support of the  German Science Foundation (DFG) Research Unit FOR2544 
`Blue Planets around Red Stars', project JE 701/3-1, and 
DFG priority program SPP 1992 `Exploring the Diversity of Extrasolar Planets (RE 1664/18)'. 
V.S. acknowledges funding from the European Research Council (ERC) under the 
European Union's Horizon 2020 research and innovation programme (grant agreement No. 682393 AWESoMeStars).
A.A.V. 
acknowledges funding received from the Irish Research Council Laureate Awards 2017/2018.

\end{acknowledgements}
\bibliographystyle{aa}
\bibliography{ref}
\begin{appendix}
\section{Surface magnetic maps}
\begin{figure*}
\centering
\subfloat[2015.77]{\includegraphics[width=0.45\columnwidth]{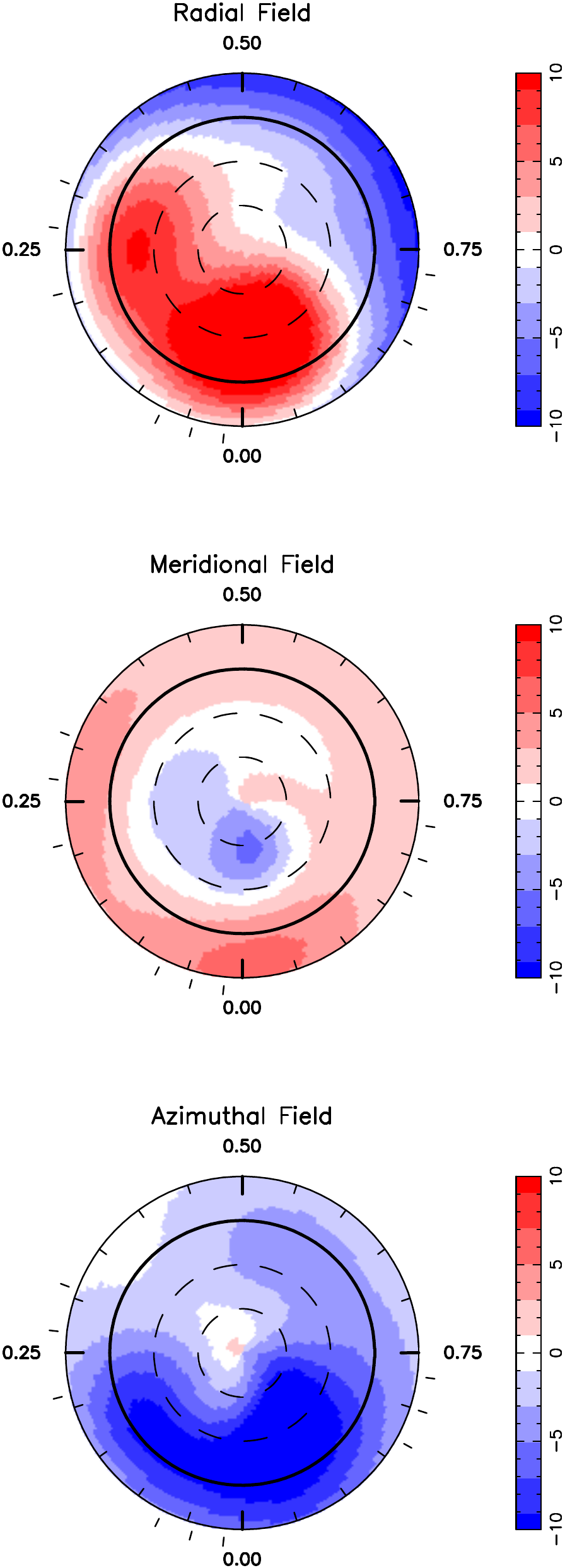}}
\qquad
\subfloat[2015.91]{\includegraphics[width=0.45\columnwidth]{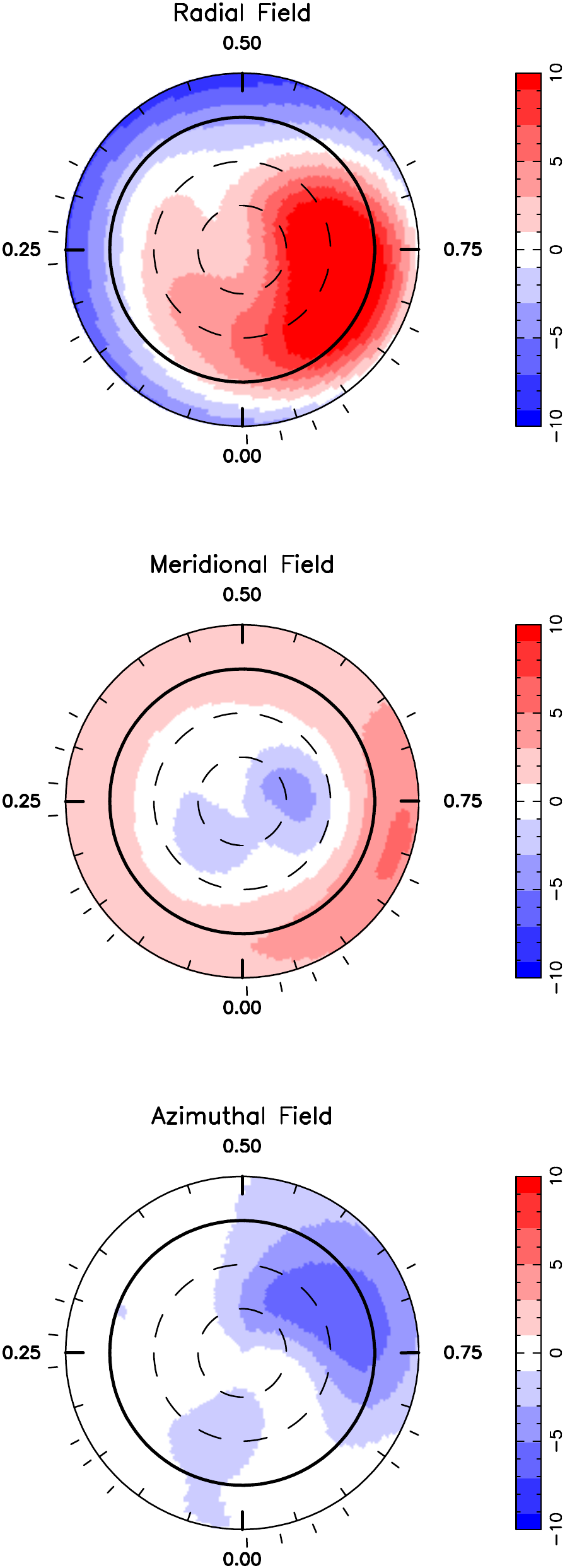}}
\qquad
\subfloat[2016.50]{\includegraphics[width=0.45\columnwidth]{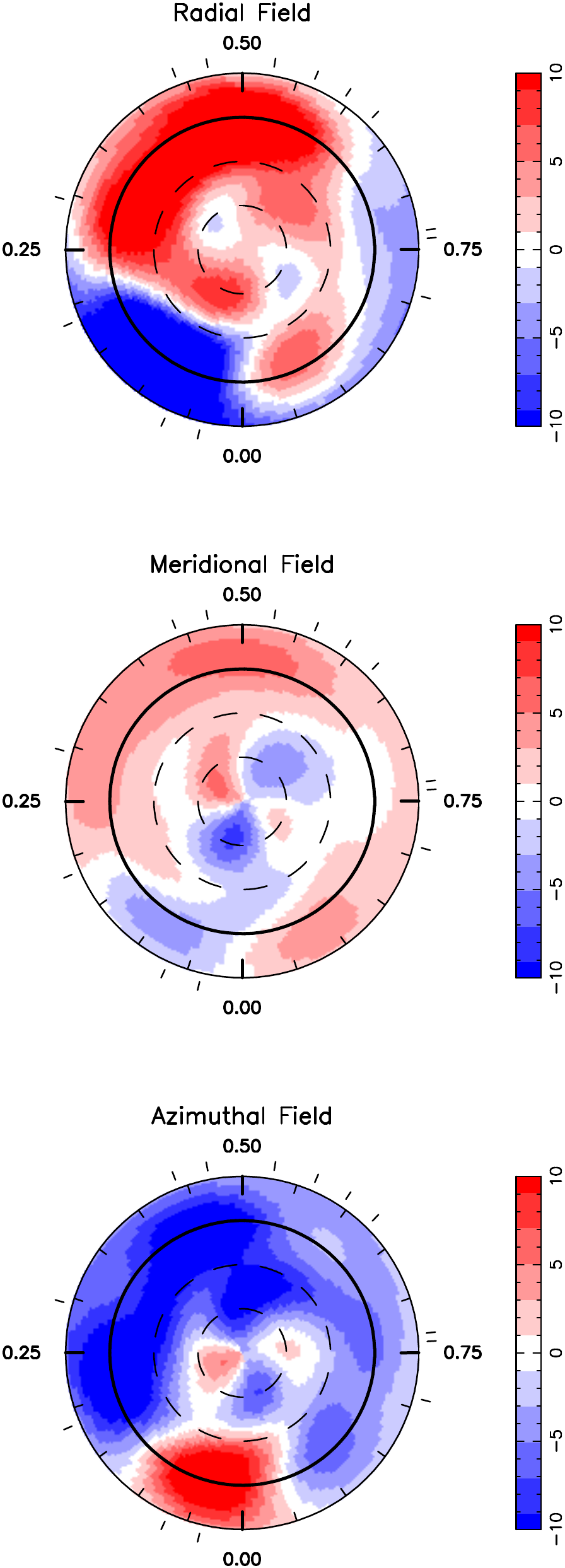}}
\caption{ZDI maps of 61 Cyg A for epochs 2015.77, 2015.91, and 2016.50 (from left to right). 
The radial field is shown at the top, the meridional field in the middle, and the 
azimuthal field at the bottom. Flatted polar projections are shown that go down to latitudes of -30$^\circ$. The
solid black line is the equator.  The small black tick marks represent the observed rotational phases. The positive and
negative fields (in gauss) are shown in  red and blue, respectively.}
\label{A1}
\end{figure*}

\begin{figure*}
\centering
\subfloat[2017.50]{\includegraphics[width=0.45\columnwidth]{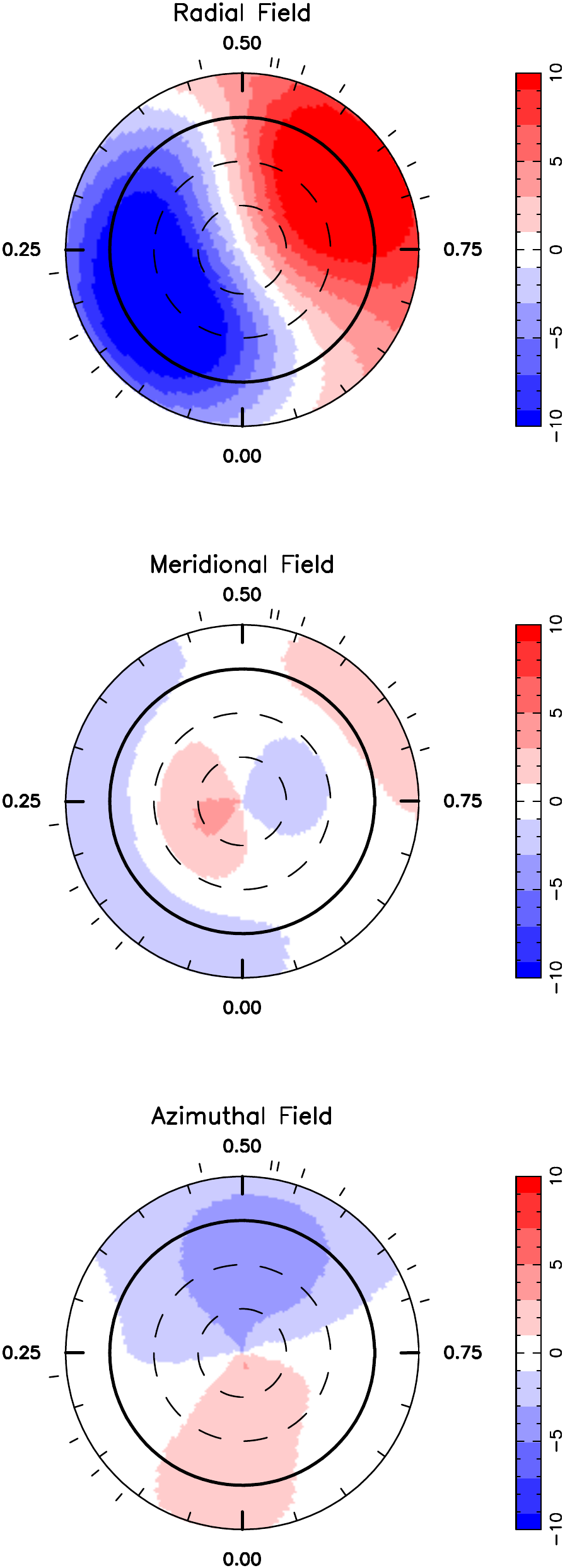}}
\qquad
\subfloat[2017.89]{\includegraphics[width=0.45\columnwidth]{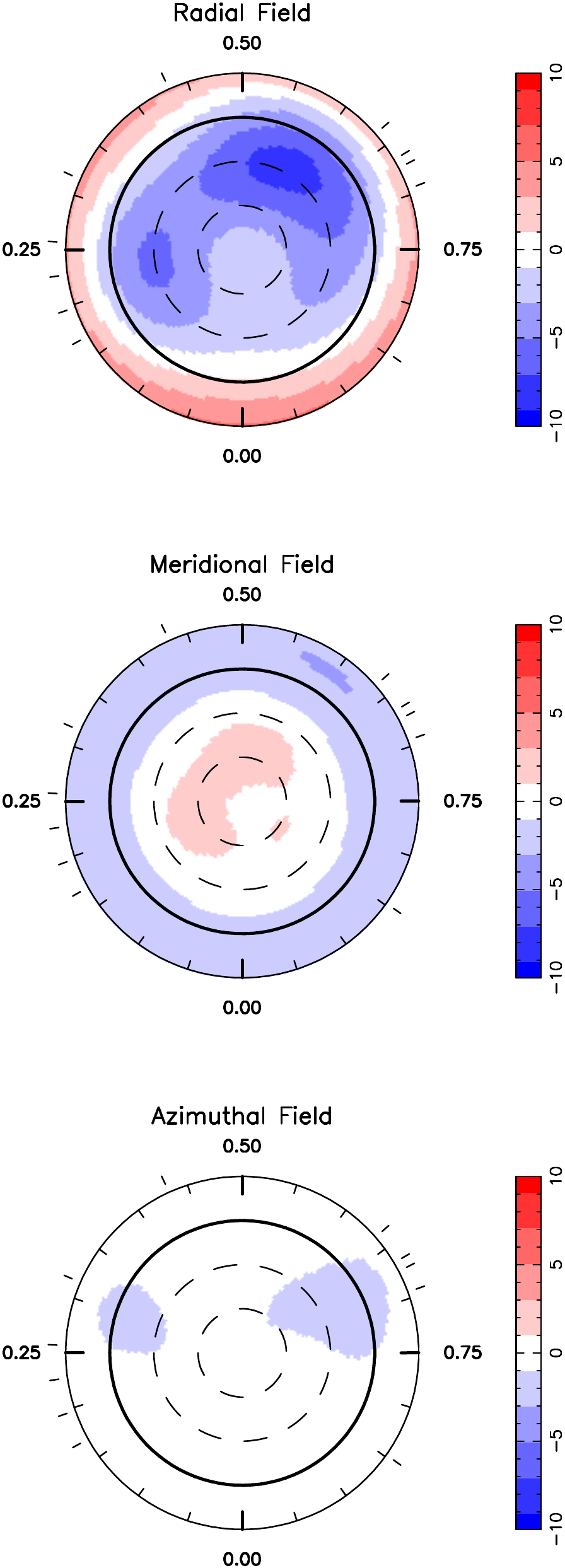}}
\qquad
\subfloat[2018.52]{\includegraphics[width=0.45\columnwidth]{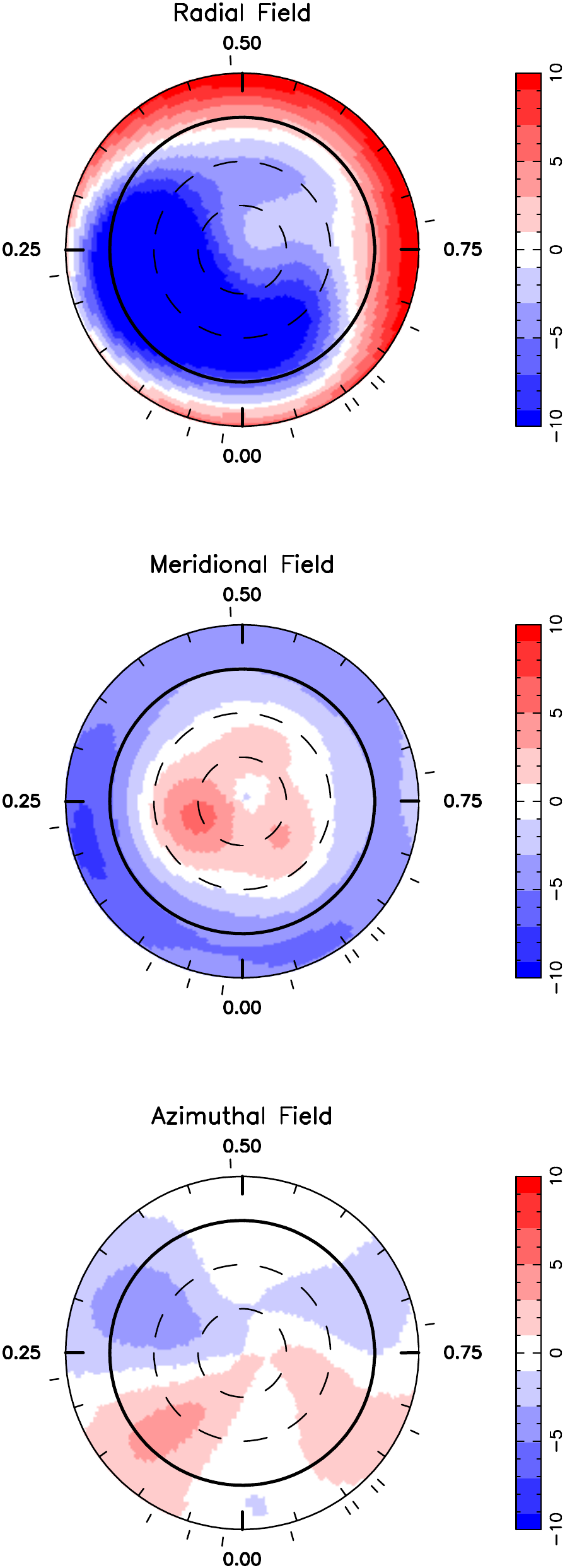}}
\caption{ZDI maps of 61 Cyg A for epochs 2017.50, 2017.89, and 2018.52. Colours and symbol as in Fig. \ref{A1}.}
\label{A2}
\end{figure*}

Figures \ref{A1} and \ref{A2} show the radial, meridional, and azimuthal field of the star during activity cycle II. 
The magnetic maps are shown in flattened polar projection.%
\paragraph{}
The first two epochs (2015.77, 2015.91) were observed within a short period of four months (see Table \ref{journal}) during intermediate chromospheric
activity. The radial field (Fig. \ref{A1}, top) 
exhibits a strong positive polarity magnetic region between the equator and higher latitudes, indicating a dipolar geometry. The overall structure of the 
field does not change during the two epochs. The meridional field also remains consistent during the two epochs with a strong band of positive polarity
field at equatorial latitudes. Contrary to the radial and meridional field, the azimuthal field has a band of negative polarity field that changes in size
within such a short period of time.
\paragraph{}
Table \ref{comps} shows that the poloidal energy changes by $\sim$20$\%$. This change in the poloidal energy could be attributed 
to the strong azimuthal field in 2015.77.  Despite a weak poloidal component in 2015.77, the dipole energy is strong. 
The axisymmetry of the field is also relatively strong during these two epochs (Table \ref{comps}). 
The increase in the strength of the azimuthal field is not reflected in the complexity of the field. 
\paragraph{}
Epochs 2016.50 and 2017.50 were taken close to activity maximum in cycle II. The S-index cycle in Fig. \ref{appendixsindex} shows that our observations 
were taken before and after the peak of  activity maximum. As the previous cycle (cycle I) had a period length of 7.2 years, 
activity maximum was not expected during these two epochs. 
Nonetheless, it is quite clear that  activity maximum occurred between 2016.50 and 2017.50. 
Both of these epochs have similar levels of activity, but the 
field geometry of 2016.50 epoch is more representative of activity maximum than epoch 2017.50. During 2016.50 the large-scale field appears 
more complex, as shown in Fig. \ref{A1}. This increase in complexity is also seen in the weakening dipole energy, followed by the 
increase in strength of higher order spherical harmonics. The strong positive polarity magnetic region from previous epochs gets smaller and a negative 
polarity region appears at equatorial latitudes. The meridional field also appears to have a more complex geometry with both positive and negative field
regions at the poles and near the equator. The azimuthal field has a dominant negative polarity with a small positive polarity region.
\paragraph{}
The large-scale field geometry changes drastically in epoch 2017.50. The complex field geometry from 2016.50 is taken over by a simple dipolar
geometry where the dipolar energy is at 92 $\%$. The radial dipole is aligned to the equator, which is shown by the 
appearance of both polarities 
in the visible hemisphere. Weak positive and negative polarities are also detected in the meridional and azimuthal field.
\paragraph{}
At maximum activity, the solar surface is dominated by 
small bipolar magnetic regions. When the solar field geometry is decomposed into its lower spherical harmonic modes, at maximum,  
it also appears to have strong 
opposite polarity magnetic regions at the equator \citep{vidotto16,vidotto18}. The radial field geometry during 2016.50 and 
2017.50 shows strong similarities with the solar large-scale field at activity maximum. 

\paragraph{}
The appearance of both positive and 
negative fields around equatorial latitudes indicates that the star's activity is changing drastically. 
The visible hemisphere will soon switch polarity from positive to negative. Unsurprisingly, we observe a strong negative polarity dipole in epoch 
2017.89. 
This transition of the large-scale dipole field is also seen in the Sun \citep{derosa12,finley18}. Recent analyses by 
\citet{vidotto18} indicate that the solar large-scale field, filtered for the large-scale components,
shows a similar trend to that presented in Figs. \ref{A1} and \ref{A2}.
\paragraph{}
During epochs 2017.89 and 2018.52 the activity decreases even further, as shown in Fig. \ref{appendixsindex}. The change in polarity appears 
before the actual activity minimum. This is not surprising as in paper I the polarity switch also occurred before the peak of 
activity minimum. 
Furthermore, the negative 
polarity magnetic field around the equator in the meridional component gets stronger with time. The azimuthal
component increases in complexity during 2018.52. This might indicate that it could change polarity during the following epochs.

\begin{figure*}
\centering
\includegraphics[width=1.\linewidth]{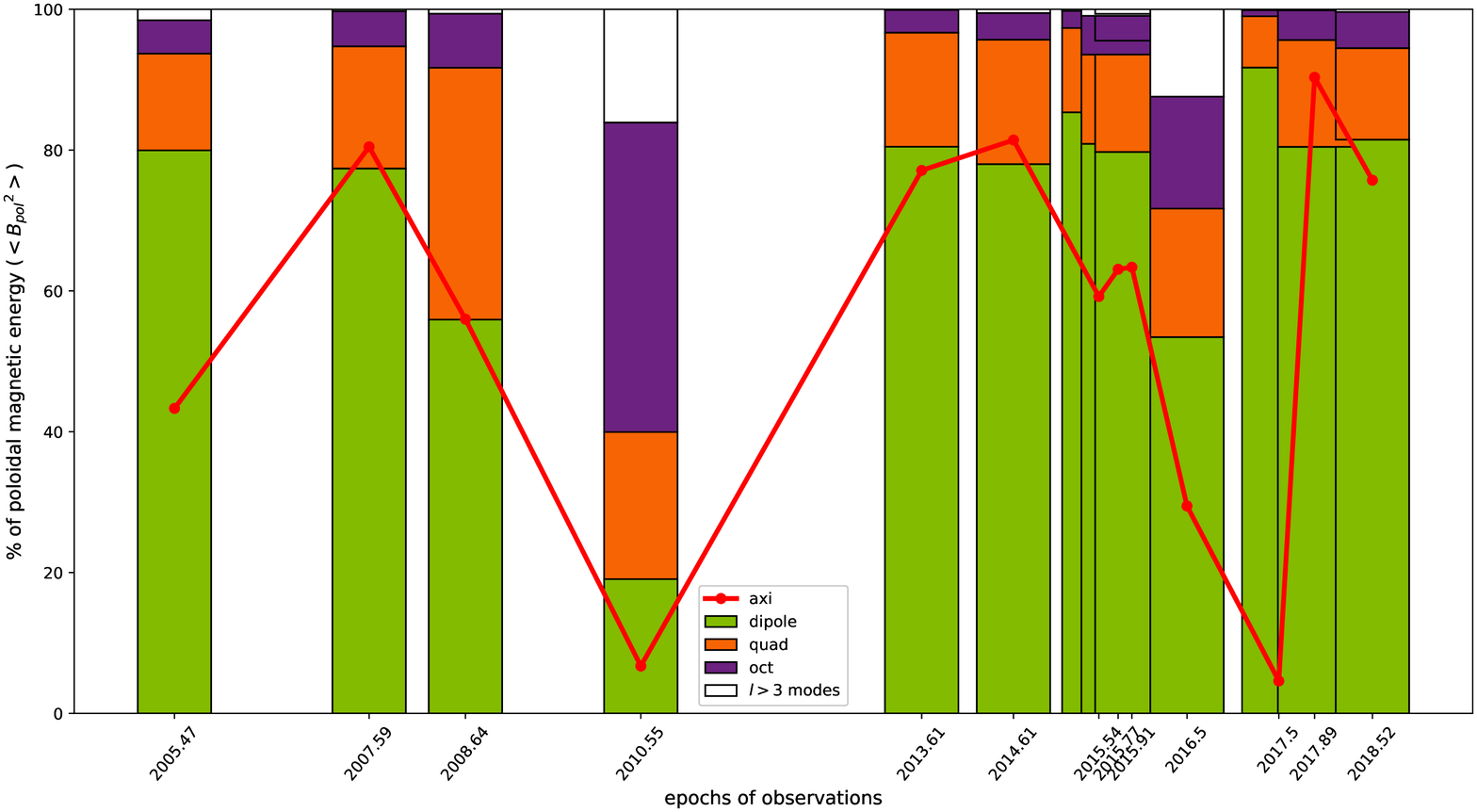}
\caption{Same as Fig. \ref{energy}, but for the entire magnetic cycle. The x-axis is scaled to time. Epochs 2007.59 to 2015.54 were previously
published in paper I.}
\label{polappendix}
\end{figure*}

\begin{table*}
\centering
\caption{Magnetic energy in the different magnetic field components. The fractional dates are shown in Col. 1, followed by the mean magnetic field (in gauss)
in Col. 2. Column 3 is the poloidal energy, Col. 4 is the poloidal dipolar energy, Col. 5 is the poloidal quadrupolar energy, Col. 6 is the 
poloidal octupolar energy, Col. 7 is the axisymmetry of the total field. The error bars are obtained using the same method as in paper I (Section 5.1). 
Finally, Col. 8 is the mean S-index and the standard deviation as error bars.}
\label{comps}
\begin{tabular}{ccccccccc}
\hline
\hline
epoch& Bmean&pol&dipole&quad&oct&axi&S-index\\
&(G)&($\%$tot)&($\%$pol)&($\%$pol)&($\%$pol)&($\%$tot)&\\
\hline
2015.77 & 9$\pm$1& 74$\pm$12& 81$\pm$10& 13$\pm$5&5$\pm$3&63$\pm$7&0.647$\pm$0.007\\
2015.91 & 6$\pm$1& 95$\pm$1& 80$\pm$6&  16$\pm$6 &  3$\pm$1 & 63$\pm$9&0.65$\pm$0.04\\
2016.50&  9$\pm$2& 79$\pm$5& 53$\pm$11&  18$\pm$4 &  15$\pm$2 & 29$\pm$6&0.67$\pm$0.01\\
2017.50 & 7$\pm$1&99${^{+1}_{-3}}$ & 92$^{+8}_{-9}$ &    7$\pm$8 & 1$^{+2}_{-1}$& 5$\pm$2&0.68$\pm$0.02\\
2017.89&4$\pm$1 & 99$\pm$1& 80$\pm$7&  15$\pm$5 & 4$\pm$2 &90$\pm$1&0.635$\pm$0.009\\
2018.52&10$\pm$2 & 99$\pm$1& 81$\pm$12&   13$\pm$4 &  5$^{+6}_{-5}$& 75$\pm$22&0.63$\pm$0.02\\
\hline
\end{tabular}
\end{table*}
\paragraph{}
Our combined observations have shown that the star's large-scale (radial and 
meridional) field changes polarity in phase with the chromospheric activity cycle. The close agreement between the radial and meridional field could
 also be attributed to cross-talk \citep{donatibrown,jeffers17}.
It takes almost twice the length of the activity cycle (one magnetic cycle) for the radial and meridional field to switch back to its original polarity.
Based on our observations, the azimuthal field switches 
polarity once during the magnetic cycle. Additionally, the large-scale field is strongly poloidal and the poloidal energy
is concentrated mostly in lower order harmonics, as shown in Fig. \ref{polappendix}. The $l_\mathrm{max}>3$ modes dominate at or near 
the two activity maxima. The poloidal field also loses its axisymmetry  close to activity maxima. The large-scale field evolution strongly indicates that
the magnetic cycle is twice the length of the activity cycle.
\section{Dipole flip as seen in Stokes $V$ profiles}
\begin{figure*}
\centering
\includegraphics[width=.75\linewidth]{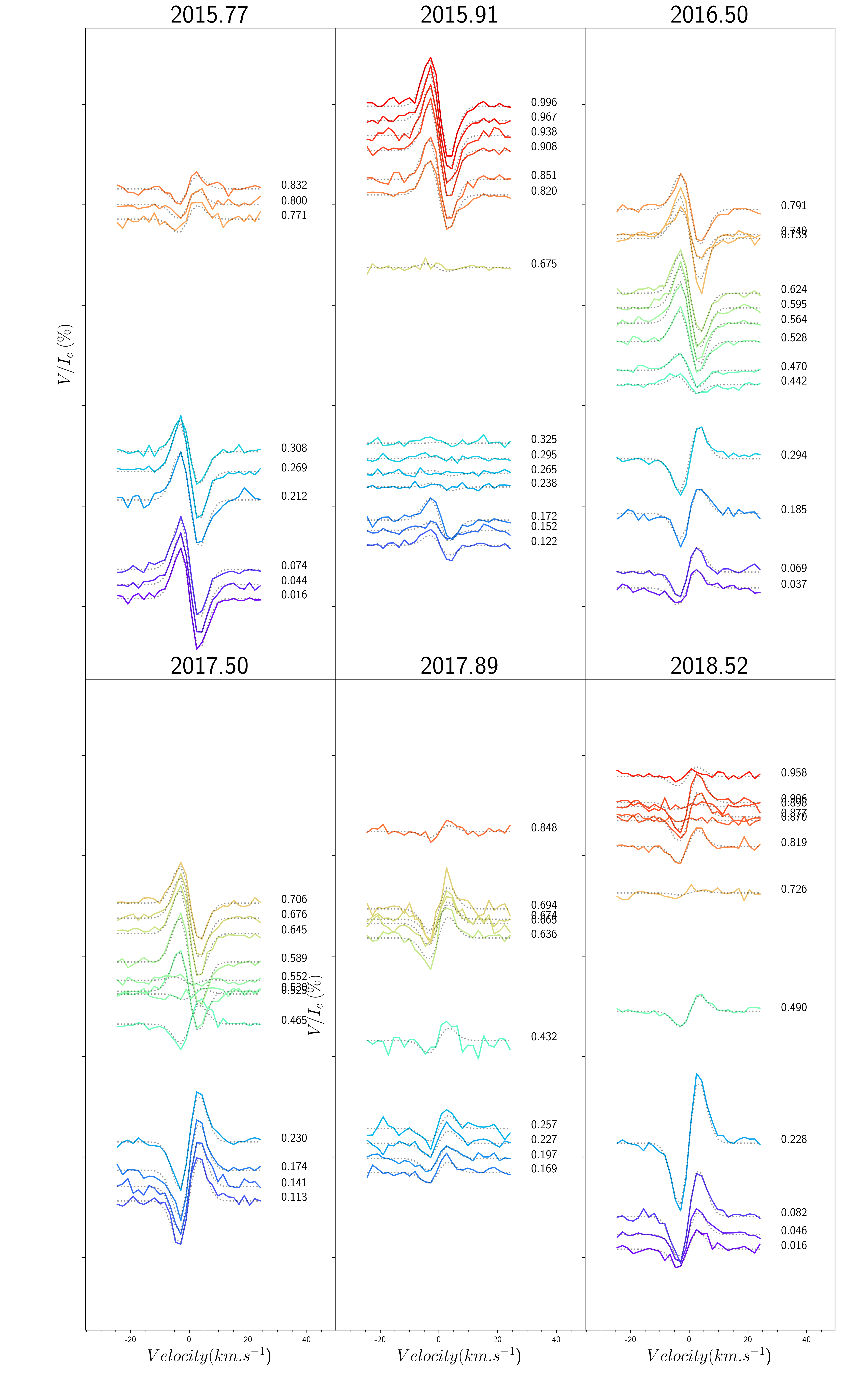}
\caption{Stokes $V$ LSD spectra of 61 Cyg A. \textit{Top}: Epochs 2015.77, 2015.91, and 2016.50. \textit{Bottom:} Epochs 2017.50, 2017.89, and
2018.52 (from $left$ to $right$). The observed profile is shown in colour and the fit is shown as dotted lines. The Stokes $V$ 
profiles are shifted according to their phase, where the phase increases from 0 to 1. The colour bar also changes with phase. The colour bar and scale of 
each subplot is kept the same.}
\label{sv}
\label{}
\end{figure*}

The LSD Stokes $V$ spectra and fits associated with the ZDI maps from Figs. \ref{A1} and \ref{A2} are shown in Fig. \ref{sv}. 
It provides a ZDI independent look at the field complexity evolution during activity cycle II. 
The amplitude of the profile indicates the magnetic field strength 
and the field polarity is determined by the sign of the profile. 
As shown in Fig. \ref{sv}, during the first two epochs in cycle II the stronger profiles all have the same sign.
Thus, it is quite clear that a single polarity dominates the visible hemisphere during these epochs. 
Although all Stokes $V$ profiles per epoch were not observed at the same 
rotational phases, the dominant Stokes $V$ profiles observed at similar phases all have the same sign, 
suggesting a dominant polarity at the visible hemisphere. As shown by the ZDI maps in Fig. \ref{A1}, the dominant polarity is positive.
\paragraph{}
Close to cycle maximum (epochs 2016.50, 2017.50) the Stokes $V$ signals exhibit mixed signs, which indicate that 
both positive and negative polarities are present and which one dominates (i.e. is the most visible) depends on rotation phase. 
The amplitudes of these oppositely aligned Stokes $V$ signals are comparably strong. 
The profiles with the same sign
are located close to each other in phase. In 2016.50, however, the profiles with a positive sign (facing right) have a better phase coverage and dominate
over the negative profiles. This is represented in the reconstructed map as a stronger positive polarity and a small negative polarity in the radial field
in epoch 2016.50.  During 2017.50 both positive and negative profiles have similar amplitude and phase coverage. This indicates that one half of the stellar 
surface is dominated by one polarity and the other half of the surface is dominated by another polarity. This is also seen in the ZDI map (Fig. \ref{A2}). 
\paragraph{} 
Finally, during the last two epochs (2017.89,2018.52) the Stokes $V$ line profiles all have the same sign, indicating a single polarity field. 
Additionally, the profiles are facing the opposite sign as the first two epochs. 
Thus, clearly the dominant magnetic field in the visible portions of the star has reversed sign. 
These two epochs have profiles that were taken during similar 
rotational phases and all of these profiles face the same sign. 
Some of the profiles in 2015.91 and 2018.52 have similar rotational phases, and as expected 
their signs are opposite. This strongly agrees with the ZDI maps in Fig. \ref{A2}.  
The LSD Stokes $V$ profiles show clear agreement with the ZDI reconstructed maps providing additional confidence to the ZDI 
technique. 

\longtab[1]{
\section{Journal of observations.}

\begin{longtable}{ccccccc}
\caption{\label{journal} Journal of observations for the six 
epochs in this work. From Left to Right: epoch, Heliocentric Julian date, Exposure time in
seconds, signal-to-noise ratio of the Stokes $V$ LSD profile, the error bar in 
Stokes $V$  LSD profile and the rotational phase.}\\
\hline\hline
Epoch& date & Julian date &Exp. time &LSD S/N &$\sigma_\mathrm{LSD}$&rotational\\
&&(2450 000+)&(s)&&10$^{-5} I_\mathrm{c}$&phase\\
\hline
\endfirsthead 
\caption{continued.}\\ 
\hline\hline 
Epoch& date & Julian date &Exp. time & LSD S/N &$\sigma_\mathrm{LSD}$&rotational\\
&&(2450 000+)&(s)&&10$^{-5} I_\mathrm{c}$&phase\\
\hline 
\endhead 
\hline 
\endfoot 
\hline
&18 September 2015&7284.43644&800&32619&3.07&0.016\\
&19 September 2015&7285.39578&800&29201&3.42&0.044\\
&20 September 2015&7286.42822&800&30034&3.33&0.074\\
&28 September 2015&7294.43000&800&30351&3.29&0.308 \\
2015.77&14 October 2015&7310.27758&800&21896&4.57&0.771 \\
&15 October 2015&7311.26826&800&30609&3.27&0.800\\
&16 October 2015&7312.33382&800&25388&3.94&0.832\\
&29 October 2015&7325.34651&800&24249&4.12&0.212\\
&31 October 2015&7327.28473&800&40144&2.49&0.269\\
\hline
&12 November 2015&7339.36547&800&33834&2.96&0.122\\
&13 November 2015&7340.39006&800&30554&3.27&0.152\\
&16 November 2015&7343.33408&800&31471&3.18&0.238\\
&17 November 2015&7344.26886&800&36146&2.77&0.265\\
&18 November 2015&7345.27477&800&36311&2.75&0.295\\
&19 November 2015&7346.32211&800&31212&3.20&0.325\\
&01 December 2015&7358.27061&800&33825&2.96&0.675\\
2015.91&06 December 2015&7363.22625&800&35305&2.83&0.820\\
&07 December 2015&7364.29177&800&25713&3.88&0.851\\
&09 December 2015&7366.25934&800&30635&3.26&0.908\\
&10 December 2015&7367.27495 &800&23688&4.22&0.938\\
&11 December 2015&7368.27652 &800&35735&2.80&0.967\\
&12 December 2015&7369.26755 &800&30829&3.24&0.996\\
&18 December 2015&7375.27907&800&35732&2.80&0.172\\
\hline
&15 May 2016&7524.55989&900&34639&2.89&0.037\\
&16 May 2016&7525.63982&900&40526&2.47&0.069\\
&20 May 2016&7529.63823&900&31859&3.14&0.185\\
&02 June 2016&7542.59157&900&34959&2.86&0.564\\
&03 June 2016&7543.63071&900&33443&2.99&0.595\\
&04 June 2016&7544.63360&900&38889&2.57&0.624\\
&08 June 2016&7548.61223&900&44345&2.26&0.740\\
2016.50&27 June 2016&7567.55244&900&39411&2.54&0.294\\
&02 July 2016&7572.59730&900&38673&2.59&0.442\\
&03 July 2016&7573.58564&900&45863&2.18&0.470\\
&05 July 2016&7575.54213&900&40408&2.47&0.528\\
&12 July 2016&7582.56750&900&22195&4.51&0.733\\
&14 July 2016&7584.53737&900&39681&2.52&0.791\\
\hline
&12 June 2017&7917.63192&800&33537&2.98&0.530\\
&14 June 2017&7919.62389&800&32305&3.09&0.589\\
&16 June 2017&7921.55409&800&34387&2.91&0.645\\
&17 June 2017&7922.60711&800&34018&2.94&0.676\\
&18 June 2017&7923.63342&800&34056&2.94&0.706\\
&02 July 2017&7937.54560&800&22982&4.35&0.113\\
2017.50&03 July 2017&7938.51815&800&26417&3.79&0.141\\
&04 July 2017&7939.63532&800&24883&4.02&0.174\\ 
&06 July 2017&7941.56775&800&31271&3.20&0.230\\ 
&14 July 2017&7949.59151&800&26993&3.74&0.465\\ 
&16 July 2017&7951.63701&800&21059&4.75&0.525\\ 
&17 July 2017&7952.57869&800&26928&3.71&0.552\\ 
\hline
&01 November 2017&8059.33560&900&18917&5.29&0.674\\
&07 November 2017&8065.30189&900&31161&3.21&0.848\\
&18 November 2017&8076.27407&900&26685&3.75&0.169\\
&19 November 2017&8077.23845&900&30423&3.29&0.197\\ 
&20 November 2017&8078.27494&900&23949&4.18&0.227\\ 
2017.89&21 November 2017&8079.27933&900&21370&4.68&0.257\\ 
&27 November 2017&8085.27097&900&14496&6.89&0.432\\
&04 December 2017&8092.25630&900&20306&4.93&0.636\\
&05 December 2017&8093.23727&900&18924&5.28&0.665\\ 
&06 December 2017&8094.23619&900&16058&6.23&0.694\\ 
\hline
&16 June 2018&8286.59636&1200&33472&2.99&0.819\\
&18 June 2018&8288.60607&1200&29781&3.36&0.877\\
&19 June 2018&8289.59103&1200&39606&2.52&0.906\\
&25 June 2018&8295.59546&1200&31257&3.20&0.082\\
&30 June 2018&8300.58423&1200&35162&2.84&0.228\\
&09 July 2018&8309.56470&1200&40728&2.45&0.490\\
2018.52&17 July 2018&8317.62340&1200&34940&2.86&0.726\\
&22 July 2018&8322.54015&1200&32160&3.11&0.870\\
&23 July 2018&8323.51315&1200&24262&4.12&0.898\\
&25 July 2018&8325.55100&1200&28374&3.52&0.958\\
&27 July 2018&8327.56221&1200&28379&3.52&0.016\\
&28 July 2018&8328.56278&1200&34145&2.93&0.046\\
\hline
\hline
\end{longtable}}
\end{appendix}
\end{document}